\documentclass[rnote]{aa} 

\usepackage{graphicx}
\usepackage{txfonts}
\usepackage{natbib}

\newcommand{\msun}{\ensuremath{{\mathrm{M}_{\sun}}}}

\begin{document}

   \title{Post-AGB stars in the 
Magellanic Clouds and neutron-capture processes in AGB stars}

   \author{M. Lugaro\inst{1,2}
          \and
          S. W. Campbell\inst{3,2}
          \and
          H. Van Winckel\inst{4}
          \and
          K. De Smedt\inst{4}
          \and
          A. I. Karakas\inst{5}
          \and
          F. K\"appeler\inst{6}
          }

   \institute{Konkoly Observatory, Research Centre for Astronomy and Earth Sciences, 
   Hungarian Academy of Sciences, Konkoly Thege Mikl\'os \'ut 15-17, H-1121 Budapest, Hungary\\
              \email{maria.lugaro@csfk.mta.hu}
         \and
             Monash Centre for Astrophysics (MoCA), Monash University, 
              Clayton 3800, Victoria, Australia\\
             \email{simon.campbell@monash.edu}
         \and 
           Max-Planck-Institut f\"ur Astrophysik, Karl-Schwarzschild-Str. 1, D-85741 Garching bei M\"unchen, 
             Germany\\
             \email{scampbell@mpa-garching.mpg.de}
         \and
           Instituut voor Sterrenkunde, K.U. Leuven, Celestijnenlaan 200D bus 2401, 
            3001, Leuven, Belgium\\
             \email{Hans.VanWinckel@ster.kuleuven.be,Kenneth.desmedt@ster.kuleuven.be}             
         \and
           Research School of Astronomy and Astrophysics, Australian National
            University, Canberra, ACT 2611, Australia\\
             \email{Amanda.Karakas@anu.edu.au}
         \and
           Karlsruhe Institute of Technology (KIT), Campus North,
           Institute of Nuclear Physics, PO Box 3640, Karlsruhe, Germany\\
            \email{franz.kaeppeler@kit.edu}
             }

   \date{}


 
  \abstract
    {} 
   {We explore modifications to the current scenario for the 
$slow$ neutron capture process in asymptotic giant branch (AGB) 
stars to account for the Pb deficiency observed in post-AGB 
stars of low metallicity ([Fe/H] $\simeq -1.2$)
and low initial mass ($\simeq$ 1 - 1.5 \msun) in the Large and Small Magellanic Clouds.} 
   {We calculated the stellar evolution and nucleosynthesis for a 1.3 \msun~star with 
[Fe/H$]=-1.3$ and tested different amounts and distributions of protons
leading to the production of the main neutron source within the $^{13}$C-pocket 
and proton ingestion scenarios.}
   {No $s$-process models can fully reproduce the abundance patterns observed in the 
post-AGB stars. 
When the Pb production is lowered the abundances of the elements 
between Eu and Pb, such as
Er, Yb, W, and Hf, are also 
lowered to below those observed.} 
   {Neutron-capture processes with neutron densities intermediate between the $s$ 
and the $rapid$ neutron-capture processes may provide a solution to this problem and 
be a common occurrence in low-mass, low-metallicity AGB stars.}

\keywords{stars: abundances -- stars: AGB and post-AGB}

\titlerunning{Post-AGB stars in the MCs and neutron-capture processes in AGB stars}
\authorrunning{M. Lugaro et al.}
 
   \maketitle
%

\section{Introduction} \label{sec:intro}

During the past decade significant information has been gathered 
on the chemical compositions
of post-asymptotic giant branch (AGB) stars in the Milky Way. This has led to the
discovery of a class of post-AGB stars that have C/O $>1$ and 
display extreme enrichments in the
abundances of the elements heavier than Fe produced by $slow$ neutron
captures
\citep[the $s$ process,][]{vanwinckel00,reyniers03,reyniers04}.
Since AGB stars can become
C-rich and have been confirmed both theoretically and observationally as
the main stellar site for the $s$ process \citep[see, e.g.,][]{busso01},
it
is natural to interpret these post-AGB observations as the signature of the
nucleosynthesis and mixing events that occurred during the preceeding AGB
phase. These events are currently identified as: (i) the mixing of protons 
into the radiative He-rich intershell leading to the
formation of a thin region rich in the main neutron source $^{13}$C 
(the $^{13}$C ``pocket''), (ii) proton-ingestion episodes (PIEs) inside the convective 
thermal pulses (TPs), and (iii) the third dredge-up
(TDU), which carries C and $s$-process elements from the He-rich intershell
to the convective envelope and to the stellar surface. Since the details of
all these processes are very uncertain
\citep[see discussion in, e.g.,][]{busso99,herwig05,campbell08}, 
observations of post-AGB stars
provide strong observational constraints. 
Recent observations of the chemical composition of four low-metallicity
([Fe/H] from $-$1.15 to $-$1.34), 
$s$-process-rich, C-rich post-AGB stars in the Large 
(J050632, J052043, and J053250) and Small 
(J004441) Magellanic Clouds (LMC and SMC, respectively)
have provided a challenge to 
AGB $s$-process models \citep{desmedt12,vanaarle13,desmedt14}\footnote{We 
do not discuss the composition of the mildly $s$-process-enhanced 
J053253 reported by 
\citet{vanaarle13} because 
this star has been recently classified as a young stellar 
object candidate, indicating that its abundances more likely reflect
the initial composition \citep{kamath15}.}. Since we know the 
distance of these stars, from the observed luminosity 
it is possible to determine that their initial stellar mass was in the range 
1 -- 1.5 \msun. Stellar AGB models in this range of mass   
and metallicity can produce the high observed 
abundances of the $s$-process elements, such as Zr and La (1 $<$ [Zr/Fe] $<$ 2 and 
1 $<$ [La/Fe] $<$ 3), together with [Pb/La] $\simeq$ 1, if a deep TDU is 
assumed after a last TP. Instead, negative [Pb/La] 
values are observed as upper limits \citep{desmedt14}. 
Here we test different possible modifications of the current AGB $s$-process scenario
to explain the neutron-capture abundance pattern 
observed in the MC post-AGB stars. 

\section{The stellar models} \label{sec:models}

\begin{figure}
\centering
\includegraphics[scale=0.5]{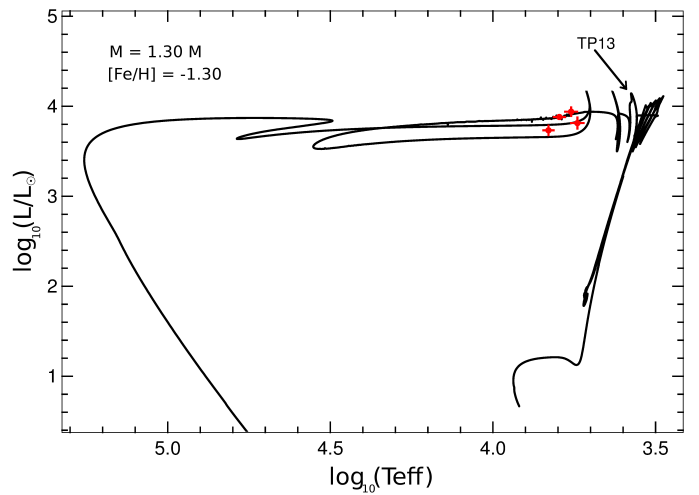}
\caption{HR diagram showing the evolution of our 1.3 \msun, [Fe/H$]=-1.3$
model from the start of the main sequence to the white dwarf cooling
track. The locations of the four post-AGB stars are represented by the red dots
with error bars. The location of the third last thermal pulse
(TP13) is also indicated. (The color version of this figure is
  available in the online journal.)}
\label{fig:HRD}
\end{figure}

We have computed the structural evolution of a star of initial mass 1.3 \msun~and [Fe/H$]=-1.3$
using the version of the Mt Stromlo/Monash evolutionary code updated by
\citet{campbell08}. Low temperature opacities have been further updated to
those calculated by \cite{lederer09}. 
Mass loss was
included using the formula of \citet{reimers75} during
the RGB phase (with $\eta = 0.4$) and the formula of \citet{vw93}
during the AGB phase. 
Instantaneous mixing was used in convective zones and the
convective boundaries were defined using a search for ``convective
neutrality'' \citep{frost96}. No overshoot was applied
beyond this, which resulted in a small
number of TDU episodes: 5 in total - from the 8$^{\rm th}$ TP to the 12$^{\rm th}$ TP --
out of a total of 15 TPs, with the amount of mass carried to the envelope, M$_{\rm TDU}$, 
of the order of a few $10^{-3}$ \msun.
The last three TPs
(13, 14, and 15) occurred after the star had left the AGB
track (Fig.~\ref{fig:HRD}). During the 15$^{\rm th}$ TP 
the post-AGB star became a {\it born-again} AGB star.
In our model this TP did not 
experience the TDU nor the ingestion of the entire H-rich envelope of mass $\sim$ 0.002 \msun.  
The observed post-AGB stars did not experience such ingestion either since they 
are not H deficient.

We computed detailed 
nucleosynthesis models using a post-processing code \citep{cannon93}
with a network of 320 nuclear species from neutrons
to $^{210}$Po. Nuclear reaction rates were from the JINA database as
of May 2012.
The initial abundances in the nucleosynthesis models were taken from
\citet{asplund09} scaled to the required [Fe/H], except for O, which was
$\alpha$-enhanced such that [O/Fe]$=+0.4$ both in the evolutionary and in the 
nucleosynthetic calculations. 
During the post-processing we artificially introduced a given abundance Y$_{\rm p}$ of 
protons at the deepest extent of each TDU episode, 
which led to the formation of the $^{13}$C pocket. 
Table~\ref{tab:TDUpocket} shows the main features of the $^{13}$C-pocket models
where Y$_{\rm p}$ is the proton abundance, M$_{\rm PMZ}$ the mass extent of 
the p-rich region, and M$_{\rm p}^{\rm tot}$ the total integrated mass of protons.
We run all of the models twice: with protons inserted after each of the five TDU
episodes, and with protons inserted only
after the last TDU episode (corresponding to the 12$^{\rm th}$ TP), except 
for pocket\_case2 which was run with protons inserted only
after the last TDU episode. 
As done by \cite{desmedt12}, we assumed a further last TDU episode to artificially occur 
in all our models after the 
13$^{\rm th}$ TP (when the mass of the envelope is
0.033 \msun). For this TDU episode we set M$_{\rm TDU}$ to the values given in 
Table~\ref{tab:TDUpocket}, such that
the observed [La/Fe] is matched for each of the four stars.
The TDU is very uncertain as it is affected by 
both physical and numerical choices \citep[see, e.g.,][]{frost96,mowlavi99}, however,
the amount of TDU does not affect the abundance patterns discussed below but only alters the 
absolute abundances.

\begin{table*}
\caption{Properties of the $^{13}$C-pocket models. M$_{\rm TDU}$ is given for the models that 
include only one $^{13}$C pocket.}
\label{tab:TDUpocket} 
\centering                          
\begin{tabular}{c c c c c c c c c} 
\hline
\hline  
 & Y$_{\rm p}$ & $^{13}$C burning$^{a}$ & M$_{\rm PMZ}$(\msun) & M$_{\rm p}^{\rm tot}$(\msun) 
 & \multicolumn{4}{c}{artificial M$_{\rm TDU}$(10$^{-3}$ \msun)} \\ 
\hline  
 &    &         &   &  & J004441 & J050632 & J052043 & J053250 \\ 
pocket\_case1 & $standard^{b}$ & radiative & $1.8 \times 10^{-3}$ & $1.35 \times 10^{-4}$ & 45.3 & 1.05 & 3.13 & 4.16 \\ 
pocket\_case2 & $standard^{b}$ & convective & $1.0 \times 10^{-3}$ & $7.50 \times 10^{-5}$  & 3.77 & 0.18 & 0.51 & 0.67 \\
pocket\_case3 & $0.70 \times 10^{-4}$ & radiative & $1.8 \times 10^{-3}$ & $1.26 \times 10^{-7}$ & 87 & 1.33 & 4.03 & 5.40 \\
pocket\_case4 & $1.05 \times 10^{-4}$ & radiative & $1.8 \times 10^{-3}$ & $1.89 \times 10^{-7}$ & 6.04 & 0.27 & 0.78 & 1.02 \\
pocket\_case5 & $1.05 \times 10^{-4}$ & radiative & $3.6 \times 10^{-3}$ & $3.78 \times 10^{-7}$ & 2.76 & 0.14 & 0.39 & 0.50 \\
\hline
\end{tabular}
\tablefoot{$^{a}$Refers to the usual case where all the $^{13}$C burns in radiative conditions during the 
interpulse period or the case where we forced the $^{13}$C to burn 
in convective conditions while it is ingested in the following TP.
$^{b}$Refers to a proton abundance Y$_{\rm p}$ that decreases exponentially with 
the mass depth from the envelope
value to $10^{-4}$ at a mass depth M$_{\rm PMZ}$ below the base of the envelope 
(Fig.~\ref{fig:profiles}).}
\end{table*}

\begin{figure*}
\centering
\includegraphics[scale=0.5,angle=270]{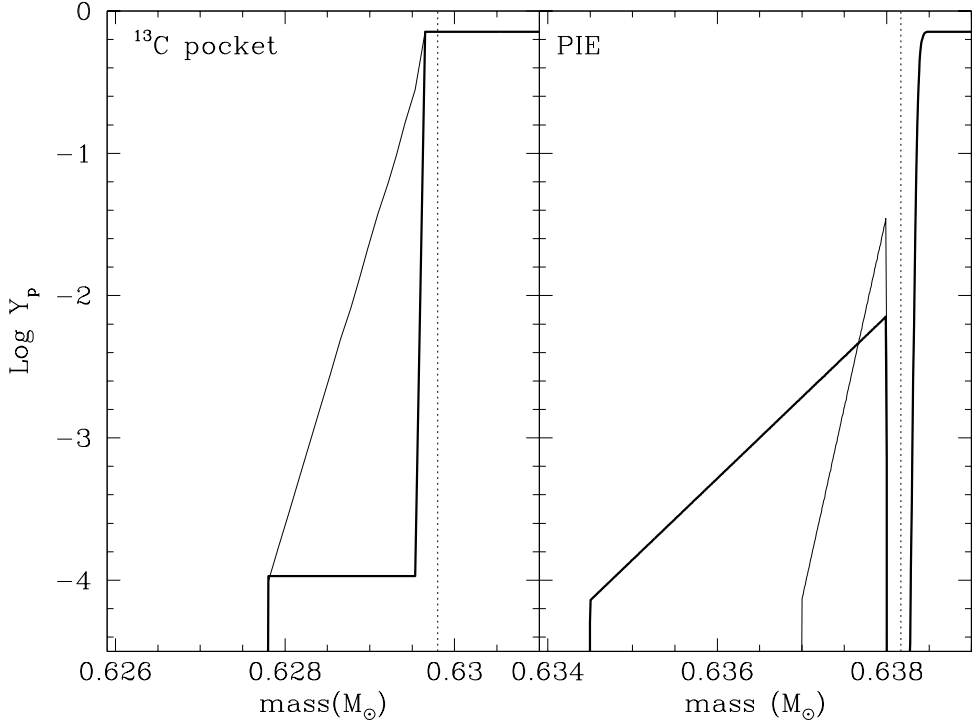} \caption{Proton abundance
distribution artificially inserted during the post-processing in our models to simulate the formation
of a $^{13}$C pocket after the 12$^{\rm th}$ TP (left panel) and a PIE event during TP13 (right panel).
{\it Left panel:} The thin line represents
the standard exponentially decreasing proton distribution (pocket\_case1)
and the thick line the case where the proton
abundance Y$_{\rm p}$ is set to constant equal $1.05 \times 10^{-4}$ (pocket\_case4).
The dotted line represents the depth
reached by the TDU.
{\it Right
panel:} The thin and thick lines represent two different proton distributions with a similar
total amount of protons ingested: PIE\_case1 (thin line) and PIE\_case2 (thick line).
The dotted line represents the maximum
extention of the intershell convective region associated with TP13. 
The steep proton profile left over by the H-burning shell is also shown in the plot
just above 0.638 \msun.}
\label{fig:profiles}
\end{figure*}

The proton abundance distribution used in pocket\_case1 (thin line of left panel in Fig.~\ref{fig:profiles}) 
is the same or very 
similar to the choice made in the models presented by \citet{desmedt12,desmedt14} 
and, as expected, produces the same results. According to our stellar evolution 
model the $^{13}$C always burns completely before 
the onset of the following TP. In pocket\_case2 we experimented with artificially 
keeping the temperature below 70 MK 
in the pocket, which 
is not enough to activate the $^{13}$C($\alpha$,n)$^{16}$O reaction.
A much lower value of the rate of 
this reaction is excluded since it is known 
within a factor of $\sim$4 \citep{bisterzo15}, but 
it is possible that a TDU episode leading to the formation of the $^{13}$C pocket occurred 
in the earlier pulses, when the core mass and the temperature are lower. 
In fact, the core mass at which the TDU begins is uncertain and may be 
lower than in our model \citep{kamath12}.
For example, if the first TDU episode occurred after the 2$^{\rm nd}$ 
instead of the 8$^{\rm th}$ TP
the temperature in the region of the $^{13}$C pocket would not have exceeed 70 MK
before the development of the next TP.
In this case the neutrons are released inside the 
following TP as the $^{13}$C is ingested. 
Next, we set the proton abundance Y$_{\rm p}$ to be  
constant with values equal to $0.70 \times 10^{-4}$ (pocket\_case3) or
$1.05 \times 10^{-4}$ (pocket\_case4, thick line of left panel of Fig.~\ref{fig:profiles}, 
and pocket\_case5). 
Physically, this may correspond to efficient mixing inside the $^{13}$C pocket during the 
interpulse period induced by the difference in the angular momentum 
between the core and the envelope in a rotating AGB star. This mixing carry 
the neutron 
poison $^{14}$N into the $^{13}$C-rich layers 
\citep[e.g.,][]{piersanti13}, lowering the $s$-process efficiency. 
However, the actual effect is uncertain because magnetic fields can slow down the core by 
coupling it to the envelope, and inhibit mixing.
In the radiative $^{13}$C-pocket pocket\_case4 the neutron flux lasts for 
roughly 30,000 yr, the  
total time-integrated neutron flux (neutron exposure) 
is $\simeq$0.5 mbarn$^{-1}$, and 
the neutron density reaches a maximum of $5.2 \times 10^6$ 
cm$^{-3}$. In the case of the convective $^{13}$C pocket (pocket\_case2)
the neutron flux lasts for a much shorter time, of the order of 10 yr, 
and the neutron density reaches a much higher maximum of $2.5 \times 10^{11}$ 
cm$^{-3}$ for a similar neutron exposure. 

\begin{table*}
\caption{Properties of the PIE models.}
\label{tab:TDUingestion}
\centering
\begin{tabular}{c c c c c c c c}
\hline
\hline
& Y$_{\rm p}^{\rm max}$ & M$_{\rm ext}^a$(\msun) & M$_{\rm p}^{\rm tot}$(\msun)
& \multicolumn{4}{c}{M$_{\rm TDU}$($10^{-3}$ \msun)} \\
\hline
& & & & J004441 & J050632 & J052043 & J053250 \\
PIE\_case1 & 0.035 & $8.0 \times 10^{-4}$ & $4.48 \times 10^{-6}$ & 0.68 & 0.03 & 0.10 & 0.13 \\
PIE\_case2 &0.007 & $3.5 \times 10^{-3}$ & $5.25 \times 10^{-6}$ & 1.08 & 0.06 & 0.16 & 0.20 \\
PIE\_case3 & 0.007 & $4.0 \times 10^{-3}$ & $6.0 \times 10^{-6}$ & 0.41 & 0.02 & 0.06 & 0.08 \\
\hline
\end{tabular}
\end{table*}

We also run some preliminary tests to simulate a proton-ingestion episode (PIE) 
during the post-processing by inserting an artificial proton abundance distribution in
the intershell before the convective zone of TP13 reaches its maximum extension.
At this time the H shell is completely extinguished. These protons are
ingested as the convective region extends outward.
As in the $^{13}$C-pocket models, we artificially
simulated a single last TDU event mixing the material from the He intershell into the envelope after TP13
had subsided, setting the amount of TDU mass to the value required to match the observed [La/Fe].
Table~\ref{tab:TDUingestion} shows the main features of the PIE models, 
where Y$_{\rm p}^{\rm max}$ is 
the maximum proton abundance, M$_{\rm ext}$ the mass extent of
the p-rich region, M$_{\rm p}^{\rm tot}$ the total integrated mass of protons, 
and M$_{\rm TDU}$ the TDU mass required to match each star.
The M$_{\rm TDU}$ required in the PIE models are typically lower than in the
$^{13}$C-pocket models.

The proton abundance distribution is set to exponentially decrease
with mass from a value Y$_{\rm p}^{\rm max} = 0.035$ (PIE\_case1, thin line of right panel of
Fig.~\ref{fig:profiles}) or 0.007 (PIE\_case2, thick line of right panel of
Fig.~\ref{fig:profiles} and PIE\_case3) to a value of 10$^{-4}$ over
M$_{\rm ext}$.
The protons are ingested in the convective region and captured
by $^{12}$C, which results in the production of $^{13}$C and subsequently
of free neutrons
for the $s$ process via $^{13}$C($\alpha,$n)$^{16}$O. We ran a series of models varying the total
mass of protons ingested M$_{\rm p}^{\rm tot}$ 
and found best solutions for M$_{\rm p}^{\rm tot} =
4.5 - 6 \times 10^{-6}$ \msun~,i.e., M$_{\rm ext} = 0.8 - 4 \times 10^{-3}$ \msun, depending
on Y$_{\rm p}^{\rm max}$. This mass extent is relatively large: a more realistic profile to
simulate overshoot of the convective intershell region into the proton tail left over by
the H-burning shell would have a higher Y$_{\rm p}^{\rm max}$  (e.g, $\sim 0.2$) and
a much smaller M$_{\rm ext}$ ($\sim 10^{-5}$ \msun, see Fig.~\ref{fig:profiles}).
However, to set and follow such a profile appropriately requires very high temporal and spatial
resolution, leading to extremely long running times
with our computational tools. We leave this task to future work.

In comparison to the scenario of the $^{13}$C pocket, the neutron flux lasts for a very
short time, of the order of 10$^{-1}$ yr, five orders of magnitudes shorter than in the
radiative $^{13}$C pocket, while the neutron density reaches a
maximum of $5 \times 10^{13}$ cm$^{-3}$ (e.g., in the PIE\_case2 model), 
seven orders of magnitude higher than in the
radiative $^{13}$C pocket. Even at this value of the
neutron density, the neutron-capture path does not move
further away from the valley of $\beta$ stability by more than two or three atomic mass numbers,
for each given element, and still proceeds sequentially
throughout each value of the proton number.

Finally, because the PIE was included in the post-processing calculations only, we
cannot evaluate its feedback on the stellar structure and evolution. 
PIEs can lead, for example, to a splitting
of the convective region, as energy released by H burning creates a temperature inversion inside the
convection zone. The amount of protons ingested in our model is small (up to $6 \times 10^{-6}$~\msun)
but this has been found in some PIE models to be enough to start the splitting of the convective zone
\citep{miller06,herwig11}. However, this effect is model dependent, e.g., the 3D PIE models of
\citet{stancliffe11} do not find the split to occur even for larger amounts of protons, while the
3D PIE models of \citet{woodward15} do. The method we have
used to simulate a PIE is clearly artificial and preliminary, 
but any 1D model not considering hydrodynamics is
necessarily inaccurate, and also 3D PIE models are not in agreement with
each other. Nevertheless, the usefulness of our approach lies in the opportunity
of identifying major qualitative differences between
the PIE and the $^{13}$C-pocket scenarios.

\begin{figure*}
\centering
\includegraphics[angle=270,scale=0.5]{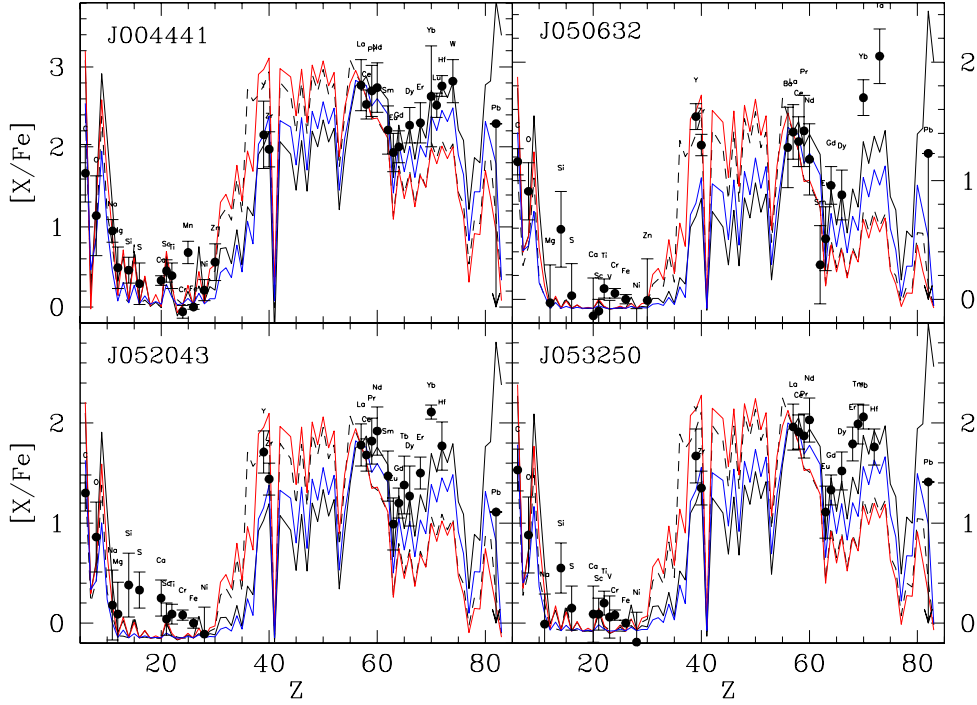}
\caption{[X/Fe] ratios as a function of the atomic number Z predicted at the stellar
  surface by the models where one $^{13}$C pocket is introduced
at the end of the TDU episode after the 12$^{\rm th}$ TP as
compared to the observations of each of the four post-AGB stars 
indicated in the panels.
We normalise both model predictions and
  observations to meteoritic rather than photospheric solar abundances for
  the elements heavier than Fe \citep{asplund09}. Notably, this modifies the 
 observed [Pb/Fe] upper limits reported by \citet{desmedt14} by $-$0.3 dex.
The black line represents pocket\_case1, which 
produces a similar 
pattern as that presented by \citet{desmedt14} with a corresponding high Pb 
abundance. pocket\_case2 is shown as a black dashed 
line. The colored lines 
represent pocket\_case3 (red) and pocket\_case4 (blue).
(The color version of this figure is available in the online journal.)}
\label{fig:1pocket}
\end{figure*}

\begin{figure*}
\centering
\includegraphics[angle=270,scale=0.5]{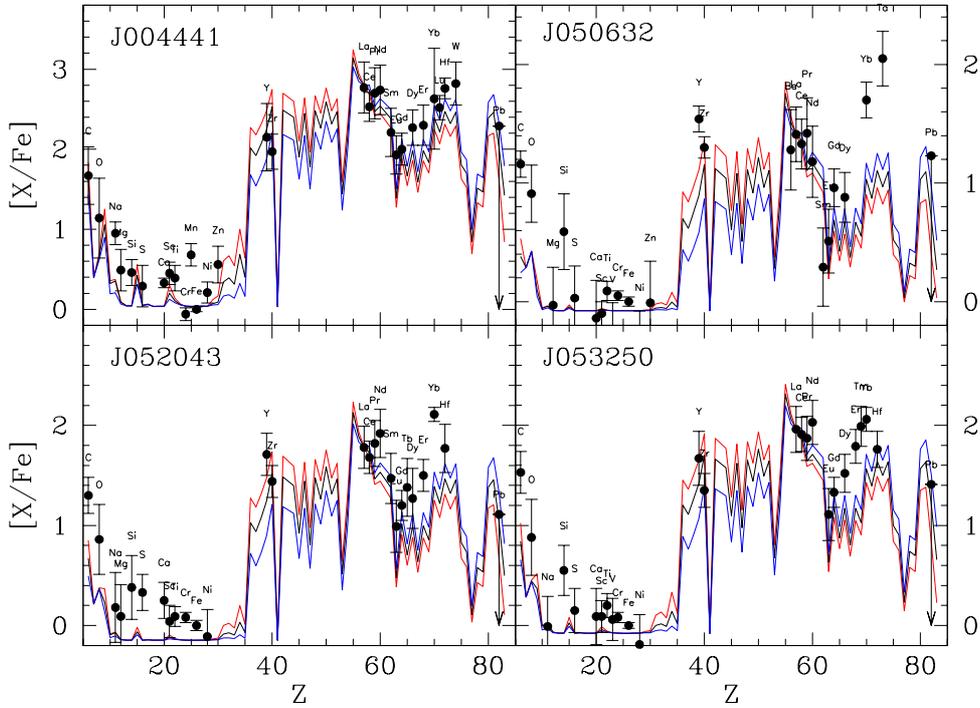}
\caption{Same as Fig.~\ref{fig:1pocket} but for the PIE models.
The black line represents PIE\_case1 (Table~\ref{tab:TDUingestion}) and
the two colored lines represent PIE\_case2 (red) and PIE\_case3 (blue).
(The color version of this figure is
  available in the online journal.)}
\label{fig:ingestion}
\end{figure*}

\section{Results} \label{sec:results}

If protons are introduced after each of the five TDU episodes computed 
by the stellar evolution code it is not possible to find a solution for any of the four stars 
since the predicted [Pb/Fe] for the models that match [Zr/Fe] is always above the observed 
upper limit (by +0.2 dex for J050632 and by +0.6 for the other three stars). 
When we include only one 
$^{13}$C pocket, instead, the production 
of Pb is lower and it is possible to find a case that simultaneously matches
[Zr/Fe] and [Pb/Fe] (Fig.~\ref{fig:1pocket}). The reason for this 
difference is that with more $^{13}$C pockets the material is recurrently 
exposed to several neutron fluxes, which pushes the abundance pattern towards Pb. 
In pocket\_case2 
M$_{\rm PMZ}$ controls the total amount of $^{13}$C ingested and consequently the neutron exposure. 
The best fit
to the observations was found using M$_{\rm PMZ} = 10^{-3}$ \msun. A main difference 
in the results between 
the {\it convective} and the {\it radiative} models is 
that the [Rb/Fe] ratio (at Z=37) is similar to [Zr/Fe] in the convective case 
while it is 1 dex lower than [Zr/Fe] 
in the radiative case. This is due to the higher neutron
density, which favours the activation of the branching point at $^{86}$Rb
producing the neutron magic $^{87}$Rb \citep{vanraai12,karakas12}.
The PIE model results are shown in Fig.~\ref{fig:ingestion}. Because the general $s$-process
elemental abundance pattern is a one-to-one function of the neutron exposure, it is possible to obtain very
similar results in the PIE models to the $^{13}$C-pocket models by setting the amount of ingested
protons to obtain a similar amount of free neutrons. As in the case of convective $^{13}$C pocket,
[Rb/Fe] is close to [Zr/Fe] in the PIE cases due to the high neutron
density. Observations of this element 
would be a powerful tool to distinguish between the different scenarios,
however, it is difficult to estimate its abundance in these stars 
because it presents only one very strong line, which is saturated.

The $^{13}$C-pocket models typically 
predict higher [C/Fe] ratios than observed, 
however, this result depends on M$_{\rm PMZ}$.
For example, in pocket\_case5 we doubled M$_{\rm PMZ}$ 
and matched the observed [C/Fe], within the error bars, for the three LMC stars.
To match the [C/Fe] in J004441, instead, requires a tenfold increase of the mass 
of the $^{13}$C pocket, 
which would cover the whole intershell. Extra-mixing 
phenomena at the base of the envelope during the red giant and AGB phases could remove C 
to produce N. Using the upper limit of the observed C abundance, the results for pocket\_case5,
and assuming that the excess C is converted into N via extra mixing, we expect [N/Fe] 
in J004441 to be greater than 2.5 dex. This is two orders of magnitude larger than 
the value of $0.61 \pm 0.5$ derived by \citet{desmedt14}. 
The PIE models, instead, typically
predict lower [C/Fe] ratios than observed, due to the lower M$_{\rm TDU}$.
The O abundances are generally underestimated by all the models, which indicates that a higher 
O abundance must be present in 
the intershell and could be achieved if the timescale of 
He burning 
during the TPs was longer or if some mixing occurred between the base of the convective zone and
the C-O core \citep{herwig00}. 
  
The four stars show positive [Y/Zr] while all the 
models predict negative values. We do not pursue this issue further because 
there may be a systematic error in the observations since the 
Y abundance is often based on a limited number of spectral lines. Another 
serious problem is that the models 
that produce [Pb/Fe] below the observational upper limits together with Y and Zr reasonably close 
to the observations underproduce the abundances of the elements between Eu and Pb, 
such as Eu, Dy, Er, Lu, Yb, Hf, and W.
The number of lines used to determine the abundances of these elements 
are limited and often blended, however, we cannot ascribe the mismatch to observational issues since 
all the four stars clearly show this problem for several of these elements.  

\section{Discussion and conclusions} \label{sec:disc+concl}

It is not possible for the $s$ process 
to produce high abundances of the elements from Eu to W together with a Pb deficiency. This 
is because once the bottleneck at the nuclei with magic number of neutrons 82 ($^{138}$Ba 
and $^{139}$La) is bypassed 
the neutron-capture flux proceeds through each isotope to $^{208}$Pb
according to its neutron-capture cross section. In other words, the abundance pattern 
between the magic neutron numbers at Ba and Pb is almost completely determined by the 
neutron-capture cross sections of the isotopes involved. These are relatively well known 
\citep{bao00} and 
that their values cannot be drastically changed is also demostrated by the fact that the $s$-process 
has no major problems (within 10-20\%) in reproducing the solar system abundances of nuclei between 
Ba and Pb that are exclusively produced 
by the $s$-process. These are $s$-only nuclei such as $^{154}$Gd, $^{160}$Dy, and $^{170}$Yb,
which typically contribute only a few percent to the total abundance of the element in the 
solar system \citep{arlandini99,bisterzo11}. 

Further observational constraints 
are available for a number of CEMP stars, which 
have metallicity roughly a factor of ten lower than the post-AGB stars considered here
and are believed to have accreted $s$-process material from 
a more massive binary companion that evolved through the AGB phase. 
\citet{bisterzo12} compared CEMP-$s$ stars (showing 
enhancements in the typical $s$-process element Ba) to AGB $s$-process models. 
They also included initial enhancements up to 2 dex in the $rapid$ 
neutron-capture process (the $r$-process) elements to  
match the Eu abundances of CEMP-$s$/$r$ stars (showing enhancements in both Ba and 
the typical $r$-process element Eu). 
The $s$-process models give a reasonable fit to the observations of all CEMP-$s$ stars 
for which at least one element between Eu and Pb is observed 
(CS\,22964-161, CS\,22880-074, CS\,29513-032, CS\,22942-019, CS\,30301-015,
CS\,30322-023, and HD\,196944). 
Four out of nine CEMP-$s$/$r$ 
stars with observations for the elements between Eu and Pb (CS\,22898-027, CS\,29497-030, 
HE\,0338-3945, and HE\,2148-1247) can also generally be matched by the models.
The remaining five CEMP-$s$/$r$ stars 
with observations for several elements between Dy and Pb  
(SDSS\,J1349-0229, CS\,31062-050, LP\,625-44, 
SDSS\,J0912+0216, and HD\,209621), instead, cannot be matched by the models because, as 
for the post-AGB stars discussed here, the observed abundances are too high even if the 
initial $r$-process abundances are enhanced. 

As suggested for CEMP-$s$/$r$ stars \citep{lugaro12}, an $s$/$r$
neutron-capture process intermediate between the $s$ and the $r$ process may have shaped 
the abundance pattern in the post-AGB stars. There are no detailed models of this 
process yet for low-metallicity AGB stars, though PIE events have been identified 
as a a possible physical site \citep[the $i$ 
process,][]{cowan77,herwig11}.
Overshoot leading to 
PIEs is uncertain, but known to be favoured in AGB stars of mass $\sim$ 1 \msun~and 
[Fe/H]$<$2
\citep{fujimoto90,campbell08,cristallo09b,lugaro12} with
recent 3D models supporting these results \citep{stancliffe11,herwig14,woodward15}. 
We stress that one-dimensional hydrostatic models of PIE events with artificial proton distributions
like those presented here are not expected
to describe the ingestion and the mixing correctly and this may have a strong effect on the
resulting abundance patterns.
However, our 1D models are still informative. For example,
we expect that the overabundance of Rb found in our PIE models may be a prominent feature
also in more accurate, e.g., 3D, PIE models \citep{herwig11}. Furthermore, the dilution
of intershell material into the envelope required
to match the observations cannot be substantially different from that we have found here; and
to avoid overproduction of Pb, also the amount of protons ingested during
a PIE cannot be substantially different.
For example, taking J053250 and comparing it with the model predictions
from the PIE model that ingested $6 \times 10^{-6}$ \msun~of protons, we could match the
observations by converting roughly 2/3 of the predicted Pb abundance into abundances of the elements
between Eu and Pb. In other words, a match to the observations would require
the same amount of free neutrons that we have in our models, but,
distributed differently among the elements between Eu and Pb. This may be possible
if the neutron density was a few orders of magnitude higher than in our models and the path
of neutron captures was shifted further away from the valley of $\beta$ stability, as in the $i$ process.

The fact that all four post-AGB stars of low 
mass considered here show similar abundance patterns in the elements heavier than Fe suggests
that the $i$ process may be a common occurrence in low-mass AGB stars up to  
[Fe/H]$\sim$1.
Since stars in this mass range are common 
this would have important implications for the stellar yields that drive  
the chemical evolution of stellar clusters and galaxies. For example, 
these long-lived low-mass stars 
may be the $i$-process source required to match observations of [Ba/Fe] and [Ba/La] 
overabundances in open 
clusters \citep{mishenina15}.
If the $i$ process is confirmed to be responsible for the abundances observed in CEMP-$s$/$r$ 
and post-AGB stars similarities and differences in the neutron-capture pattern of the 
two groups, which sample different metallicity ranges, will
provide fundamental constraints to pin down its metallicity dependence and its impact on the chemical 
evolution of stellar systems.

\begin{acknowledgements}

We thank Marco Pignatari, Carolyn Doherty for discussion and suggestions,  
and JINA for providing the online {\it reaclib} database. ML is a Momentum Project 
leader of the Hungarian Academy of Sciences. AIK is an ARC Future Fellow (FT10100475).
HVW and KDS acknowledge support of the KULeuven fund GOA/13/012.
This research was supported under the Australian Research 
Councils Discovery Projects funding scheme (project numbers DP1095368 and DP120101815) and 
by the computational resources provided by the Monash e-Research Centre via the Australian 
Research Councils Future Fellowship funding scheme (FT100100305) and in part by the Australian Government 
through the National Computational Infrastructure under the National Computational Merit Allocation Scheme 
(projects g61 and ew6).

\end{acknowledgements}


\bibliographystyle{aa}
\bibliography{apj-jour,library}

\end{document}